\documentclass[%
 aip,
 amsmath,amssymb,
 reprint,%
]{revtex4-1}

\usepackage{graphicx}
\usepackage{dcolumn}
\usepackage{bm}

\usepackage[utf8]{inputenc}
\usepackage[T1]{fontenc}
\usepackage{mathptmx}
\usepackage{etoolbox}
\usepackage{bigints}
\usepackage[dvipsnames]{xcolor}
\usepackage{comment} 

\makeatletter
\def\@email#1#2{%
 \endgroup
 \patchcmd{\titleblock@produce}
  {\frontmatter@RRAPformat}
  {\frontmatter@RRAPformat{\produce@RRAP{*#1\href{mailto:#2}{#2}}}\frontmatter@RRAPformat}
  {}{}
}%
\makeatother
\begin{document}

\preprint{AIP/123-QED}

\title[Transport properties of the square-well fluid]{Transport properties of the square-well fluid from molecular dynamics simulation}

\author{Iv\'an M. Zer\'on}
\affiliation{Laboratorio de Simulaci\'on Molecular y Qu\'imica Computacional, CIQSO-Centro de Investigaci\'on en Qu\'imica Sostenible and Departamento de Ciencias Integradas, Universidad de Huelva, 21006 Huelva Spain}

\author{Marina Cueto-Mora}
\affiliation{Laboratorio de Simulaci\'on Molecular y Qu\'imica Computacional, CIQSO-Centro de Investigaci\'on en Qu\'imica Sostenible and Departamento de Ciencias Integradas, Universidad de Huelva, 21006 Huelva Spain}

\author{Felipe J. Blas}
\affiliation{Laboratorio de Simulaci\'on Molecular y Qu\'imica Computacional, CIQSO-Centro de Investigaci\'on en Qu\'imica Sostenible and Departamento de Ciencias Integradas, Universidad de Huelva, 21006 Huelva Spain}
\email{felipe@uhu.es}

\begin{abstract}

In this work, we have calculated self-diffusion and shear viscosity, two of the most important transport properties, of the spherical square-well (SW) fluid interacting with potential range $\lambda = 1.5 \, \sigma$. To this end, we have used a combination of molecular dynamics simulation and the continuous version of the square-well (CSW) intermolecular potential recently proposed by Zer\'on \emph{et al.} [Mol. Phys. \textbf{116}, 3355 (2018)]. In addition to that, we have also determined a number of equilibrium properties, including internal energy, compressibility factor, radial distribution function, and coordination number. All properties are evaluated in a wide range of temperatures and densities, including subcritical and supercritical thermodynamic conditions. Results obtained in this work show an excellent agreement with available data reported in the literature and demonstrate that the CSW intermolecular potential can be used in molecular dynamics simulations to emulate SW transport properties with confidence.

\end{abstract}


\maketitle

%

\section{Introduction}

The spherical square-well (SW) force field is one of the simplest intermolecular potentials that incorporates both repulsive and attractive interactions. Although its mathematical expression shows discontinuities at $r=\sigma$ and $r=\lambda\sigma$, with $r$ the center-to-center distance between two interacting sites of diameter $\sigma$ and intermolecular potential range $\lambda$, it has been regularly used to describe the interactions 
between simple and real complex fluids.\cite{bolhuis1997numerical, zhou1997equilibrium, foffi2002phase,kern2003fluid,lu2008gelation,valadez2012phase,rouwhorst2020nonequilibrium,rouwhorst2020nonequilibrium2} Due to its simple mathematical form, SW fluids have been extensively investigated using perturbation theory\cite{zwanzig1954high,barker1967perturbation, smith1970approximate, smith1971perturbation, smith1974percus, henderson1976monte, smith1977mean,del1987properties, del1987properties2, benavides1989properties, benavides1991properties, chang1994completely, gil1997statistical, benavides1999thermodynamics} and molecular simulation using the well-known Monte Carlo (MC) technique.\cite{carley1983thermodynamic, benavides1991properties, heyes1992square, vega1992phase, green1994vapor, de1997critical, orkoulas1999phase, elliott1999vapor, labik1999sp, del2002vapour, orea2003surface, orea2004liquid, largo2003theory, singh2003surface, pagan2005phase, largo2005pair, scholl2005vapor, patel2005generalized, espindola2009optimized, pavlyukhin2012thermodynamic, sastre2015microcanonical, sastre2018microcanonical, sastre2021helmholtz}

Structural and equilibrium properties have been routinely evaluated using the aforementioned theoretical and numerical procedures, including radial distribution function, coordination number, internal energy, compressibility factor, coexistence densities, vapor pressure, interfacial tension, high-temperature expansion coefficients of free energies, and chemical potential, among many others. Unfortunately, there is a limited number of publications in the literature focusing on the determination of transport properties in systems that interact through discontinuous potentials from computer simulation, and the SW fluid is not an exception. In this context, the series of works published in the 80's by Michels and Trappernier is particularly interesting and relevant to this study.\cite{michels1979molecular,michels1980molecular,michels1980-2molecular,michels1981molecular,michels1981-2molecular,michels1982molecular}

The limited number of studies addressing transport properties of fluids interacting through discontinuous intermolecular potentials may be attributed to two primary factors. Firstly, the calculation of transport properties is unfeasible through Monte Carlo (MC) simulation, as it requires the continuous tracking of all particles' trajectories over time. Secondly, investigating discontinuous potentials in molecular dynamics (MD) simulations using conventional methods is not feasible, as the continuous application of forces is indispensable for solving Newton’s equations.

To address this issue, authors have previously suggested two distinct approaches. The initial option involves modifying the molecular dynamics algorithm itself. A viable choice is to adopt the collision scheme, replacing the solution of Newton's equations. In this scheme, energy and momentum are conserved quantities both before and after the collision.\cite{alder1959studies, chapela1984molecular}  Through this algorithm, direct particle contact is circumvented, enabling the computation of transport properties. This approach is exemplified in the works of Michels and Trappernier. Unfortunately, the adoption of this technique is not widespread.

The second commonly suggested solution involves conducting conventional MD simulations and adjusting the discontinuous potential to a softened version with the capability to replicate most, if not all, equilibrium and non-equilibrium properties by adjusting multiple free parameters. \cite{fomin2008quasibinary, abraham2011liquid, jover2012pseudo, zeron2018continuous,sandoval2022soft, munguia2022generalized} Examples of this approach are the soft version of the hard sphere\cite{jover2012pseudo} and the continuous version of the SW intermolecular potential (CSW) recently proposed by Zer\'on \emph{et al.}{\cite{zeron2018continuous} 

In this investigation, we adopt the latter approach. Consequently, the principal aim of this study is to determine self-diffusion, shear viscosity, and diverse structural and thermodynamic properties of the SW fluid through the integration of molecular dynamics simulations with the CSW intermolecular potential.

The organization of the paper is as follow: In Sec. II, we describe the methodology used in this work to estimate transport properties, particularly, self-diffusion coefficients and shear-viscosity using standard equilibrium MD simulations. Specific details related to the MD implementation can be found in Sec. III.  The results obtained as well as their discussion, are described in Sec. IV. Finally, conclusions are presented in Sec. V. Numerical data of this work has been included as Supplementary Material.

\section{Transport properties by computer simulations}

Transport phenomena involves the study of a number of properties in systems in which mass, energy, charge, momentum or angular momentum is exchanged. MD simulations employ two frameworks to calculate transport properties: equilibrium and non-equilibrium. In both scenarios, established formalisms and equations exist for computation. For instance, equilibrium MD simulations commonly utilize the Green-Kubo expression and the Einstein  formula.

When considering the equilibrium state of the system, it becomes possible to express a given transport property in terms of an appropriate correlation function of a dynamic magnitude. In the cases of self-diffusion and shear viscosity, the dynamic properties are linked to the positions of the particles constituting the system and the pressure tensor, respectively. Both are defined in the subsequent sections. We recommend the reader the manuscript of Maggin \emph{et al.}~\cite{maginn2019best} for a detailed account on how to calculate self-diffusivity and viscosity from equilibrium MD.

\subsection{Self-diffusion coefficient}

The most common way for determining the self-diffusion coefficient, $D$, involves the Einstein approach. In this approach, the dynamic magnitude is directly related with the positions of the particles, as functions of time, and $D$ is proportional to the slope of the mean-square displacement (MSD) at long times:

\begin{equation}
    D = \frac{1}{6}\, \lim_{t\to \infty} \,\, \frac{d}{dt} \left\langle \dfrac{1}{N}\sum_{i=1}^{N} [r_{i}(t)-r_{i}(0)]^{2} \right\rangle
    \label{eq_D}
\end{equation}
where $r_{i}(t)$ is the position of the $i$-particle at time $t$. Note that the ensemble average in Eq.~\eqref{eq_D} is taken for all $N$ particles in the simulation to improve statistical accuracy of this property.\cite{maginn2019best,Allen2017a}

\subsection{Shear viscosity}

The shear-viscosity, $\eta$, can be calculated with the well-known Green–Kubo formula:
\begin{equation}
    \eta = \frac{V}{k_{B}T} \int_{0}^{\infty} \langle P_{\alpha \beta}(t_{0}) P_{\alpha \beta}(t_{0}+t) \rangle_{t_{0}} \,dt \,,
    \label{eq_eta}
\end{equation}
where $V$ is the volume of the system, $k_{B}$ the Boltzmann constant, $T$ the temperature, $\langle P_{\alpha \beta}(t_{0}) P_{\alpha \beta}(t_{0}+t) \rangle_{t_{0}}$ is the auto-correlation function of the pressure tensor, which measure the rate at which the momentum in direction $\beta$ is transferred along the $\alpha$ direction. The components of the pressure tensor, $P_{\alpha \beta}$, can be expressed as:
\begin{equation}
    P_{\alpha \beta} = \frac{m}{V} \sum_{i=1}^{N} v_{i}^{\alpha}\,v_{i}^{\beta} + \frac{1}{V}\sum_{i=1}^{N-1}\sum_{j=i+1}^{N}r_{ij}^{\alpha}\, f_{ij}^{\beta}
\end{equation}

The first term is the kinetic contribution, being $m$ the molecular mass, $v_{i}^{\alpha}$ the $\alpha$ component of the velocity of the particle $i$, $\mathbf{v}_{i}$. The second term is the intermolecular potential contribution, where $r_{ij}^{\alpha}$ is the $\alpha$ component of the intermolecular vector $\mathbf{r}_{ij}$ between particles $i$ and $j$, and $f_{ij}^{\beta}$ is the $\beta$ component of the vector force of $j$ over $i$ particle, $\mathbf{f}_{ij}$. In homogeneous isotropic systems $P_{\alpha \beta}$ is symmetric and $\eta$ can be enhance averaging multiple terms from the stress tensor.

\section{Simulation details}

All simulations have been performed in the NVT or canonical ensemble using the
well-known GROMACS package (double precision).\cite{VanDerSpoel2005a,hess2008gromacs}
To calculate all properties, we are using argon fluids interacting with the continuous version of the SW potential proposed by Zer\'on \emph{et al.}\cite{zeron2018continuous} The reduced intermolecular potential, $v_{_{CSW}}(x) = V_{_{CSW}}(x)/\epsilon$, as a function of the reduced distance, $x=r/\sigma$, is written as:

\begin{equation}
    v_{_{CSW}} (x) = \frac{1}{2}\left(\left( \frac{1}{x} \right)^{n} + \frac{1- e^{-m(x-1)(x-\lambda)} }{1+e^{-m(x-1)(x-\lambda)}} -1  \right)
    \label{eq_csw_pot}
\end{equation}
Here, the hard sphere diameter value is $\sigma = 0.3405\,\text{nm}$, the well energy $\epsilon = 0.996078\,\text{kJ/mol}$, and the interaction range $\lambda =1.5 \, \sigma$. The softness parameters, $n=2500$ and $m=20000$, are taken from the original work.\cite{zeron2018continuous} 
The CSW potential and its derivative are input into GROMACS through a file, where numerical values are tabulated at intervals of $0.0001\,\text{nm}$.

We have performed simulations using two system sizes, $108$ and $1000$ molecules. The first election allows to compare our results obtained using the CSW intermolecular potential with simulation data taken from the literature for systems formed from the same number of molecules but interacting with the original SW potential. The second election comprises a significantly larger system. Comparison between results obtained using both system sizes will allow to analyze the possible existence of system-size effects on the transport properties calculated in this work. In both cases, the initial simulation boxes are prepared placing spherical molecules interactions via the CSW at random positions in a cubic simulation box with periodic boundary conditions applied in all directions. The studied densities range from $160$ to $1450\,\text{kg/m}^{3}$.

In all cases, we firstly run all simulations $2\,\text{ns}$ to stabilize the system, followed by $20\,\text{ns}$ runs to obtain averaged properties. The leap-frog integrator\cite{Cuendet2007a} is employed to solve the dynamic equations with a time step of $0.1\,\text{fs}$. Temperature is controlled using the Nosé-Hoover thermostat\cite{Nose1984a, Hoover1985a} with a coupling constant of $\tau=2\,\text{ps}$ in the range of temperatures studied, $80-600\,\text{K}$. The positions of all particles are saved each 50 steps and pressure components each 5 steps.

Since we are using the CSW version of the potential, GROMACS treats it as a standard intermolecular continuous potential. Consequently, a cut-off distance must be provided. In our calculations, $U_{_{CSW}}(r_{c})\approx 10^{-15}\,\text{kJ/mol}$ when $\lambda =1.5 \,\sigma$ and the cut-off distance is set equal to $r_{c}=0.512\,\text{nm}$. Since the potential value is practically equal to zero, we do not apply long-range corrections during simulations.

All the results obtained in this work are presented in reduced units: density, $\rho^{*}= \rho \, \sigma^{3}$, temperature, $T^{*} = k_{B}T/\epsilon$, internal energy per particle, $u^{*} = U/Nk_{B}T$, pressure, $P^{*}=P\sigma^{3}/\epsilon$, compressibility factor, $Z=P^{*}/(\rho^{*}T^{*})$, self-diffusion coefficient, $D^{*}= D \sqrt{m/(\sigma^{2}k_{B}T)}$, and shear-viscosity, $\eta^{*} =\eta \sigma^{2}/ \sqrt{(mk_{B}T)}$.

\section{Results}

The ability of the CSW intermolecular potential in reproducing equilibrium properties of spherical SW fluids in bulk vapor or liquid phases, as well as at vapor-liquid coexistence conditions, has been addressed in the original work by Zerón \emph{et al.}\cite{zeron2018continuous} That study has demonstrated excellent agreement between predictions obtained from MD simulations using the CSW potential and those from MC using the original SW force field. The comparison includes a number of properties, encompassing internal energies, compressibility factors, radial distribution functions, coexistence densities, vapor pressures, and vapor-liquid interfacial tensions. However, transport properties have not been previously explored using this continuous version of the intermolecular potential.

To demonstrate the suitability of the CSW intermolecular potential for standard MD simulations in accurately predicting the transport properties of the SW fluid, we 
perform simulations to determine the self-diffusion coefficient and shear viscosity of particles interacting via the CSW potential, with $\lambda=1.5\,\sigma$, under subcritical and supercritical conditions. The thermodynamic states considered in this work are illustrated in Fig.\ref{fig_datos}. Note that the majority of the chosen states align with those previously studied by Michels and Trappeniers.\cite{michels1980molecular, michels1980-2molecular, michels1982molecular}

Firstly, we compute the self-diffusion coefficients and shear viscosity for a system formed by$108$ particles in order to compare the MD results obtained using the CSW potential with values found by Michels and Trappeniers (using the standard SW potential). It is noteworthy that we employ the same system size as Michels and Trappeniers to avoid differences due to finite-size effects. Subsequently, we conduct simulations under identical thermodynamic conditions, but this time employing systems with $1000$ molecules. As we have already mentioned, this allowed us to analyze the impact of the number of molecules on both transport properties and equilibrium and structural properties.

To quantify the ability of the CSW intermolecular potential in predicting transport, equilibrium, and structural properties of the standard SW potential, we also calculate the average absolute relative deviation (AAD\%): 
\begin{equation}
    \text{AAD}\% = \frac{100}{n_{t}} \sum_{i=1}^{n_{t}} \left| \frac{X^{CSW}_{i} - X^{SW}_{i}}{X^{SW}_{i}} \right| \,,
\end{equation}
where $X^{CSW}_{i}$ and $X^{SW}_{i}$ are the values of a magnitude calculated using the CSW potential via MD simulations and using the standard SW potential with MC simulations, respectively. Here $i$ runs for all the simulation values used in the calculation and $n_{t}$ is the total number of values.

\begin{figure}
\includegraphics[width=\columnwidth]{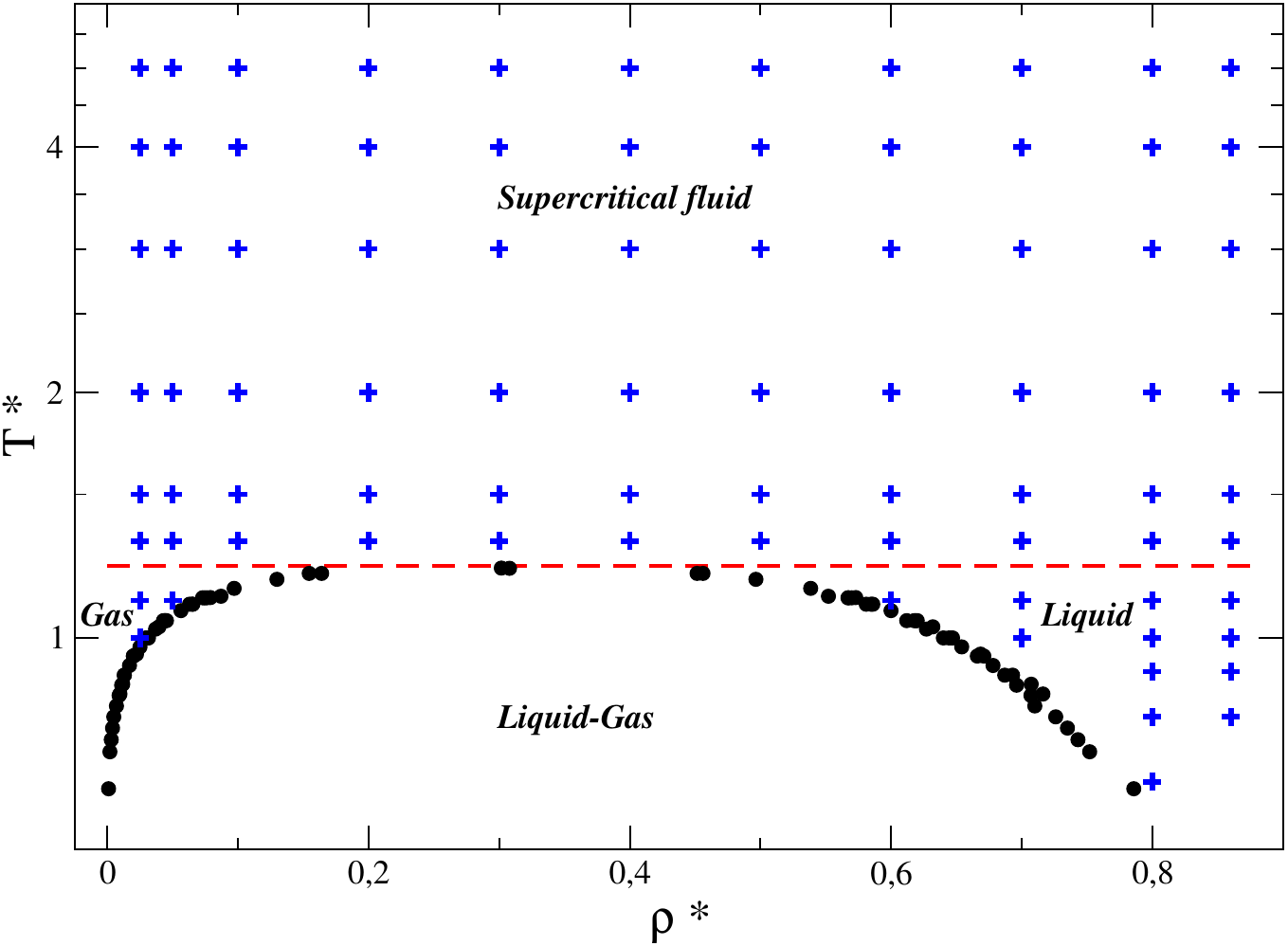}
\caption{ $T^{*}-\rho^{*}$ projection of the phase diagram of the SW fluid with range $\lambda=1.5\,\sigma$. The blue plus represent the thermodynamic states studied in this work, the black circles are the coexistence densities reported by several authors\cite{vega1992phase,green1994vapor,elliott1999vapor,del2002vapour,singh2003surface,orea2003surface,patel2005generalized} and the discontinuous red line separates the subcritical and supercritical regions of the phase diagram at the critical temperature.}
\label{fig_datos}
\end{figure}

\subsection{Self-diffusion coefficient}

We have performed MD-NVT simulations for systems formed by $108$ molecules interacting through the CSW intermolecular potential with range $\lambda =1.5 \,\sigma$. Particularly, we have simulated states at densities from $\rho^{*}=0.025$ to $0.86$ and in the range of temperatures $T^{*}=0.667-5$.
For each point studied, 5 independent simulations of $20\,\text{ns}$ are running to determine the average value and to estimate the uncertainty associated to the self-diffusivity. The numerical value has been calculated using the Einstein relation of Eq.~\eqref{eq_D}, i.e., we take one sixth of the slope of the linear fit applied to the $10-90\%$ curve of the MSD in the diffusivity zone. 

The final values of the self-diffusion coefficient of spherical SW molecules obtained from MD-NVT simulations using the CSW intermolecular potential are shown in Fig.~\ref{fig_D_108}. We have also included the values obtained by Michels and Trappeniers\cite{michels1982molecular} using the standard SW potential model. As can be seen, there is an excellent agreement between both results at all temperatures and densities ($\text{AAD\%}\approx 3\%$). It is important to remark that estimations obtained by Michels and Trappeniers were calculated using discontinuous Molecular Dynamics and the Green-Kubo formula for a system with the same size but interacting via the standard SW intermolecular potential.\cite{michels1982molecular} 

\begin{figure}
\includegraphics[width=\columnwidth]{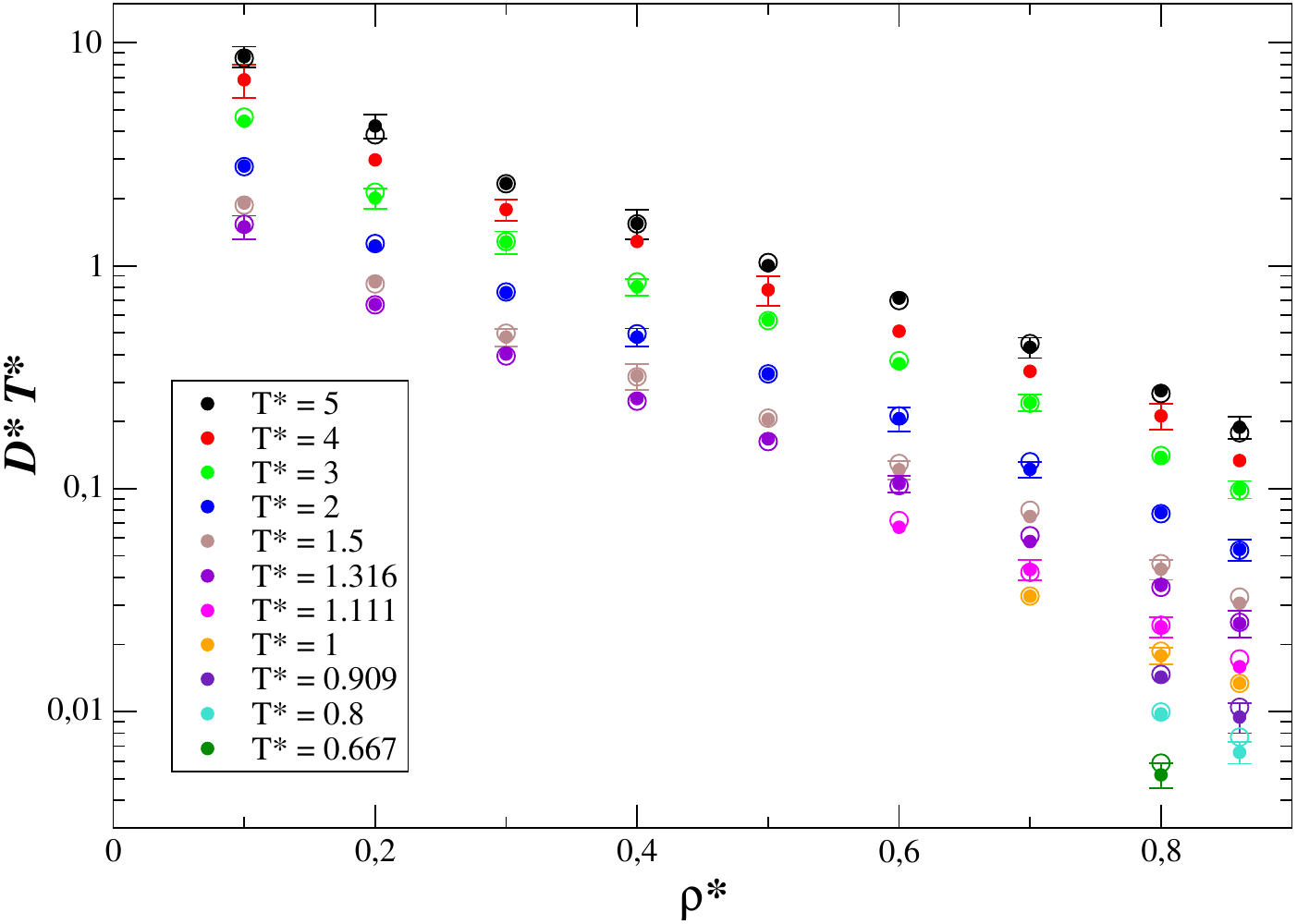}
\caption{Self-diffusion coefficient $D^{*}$, as a function of density, $\rho^{*}$, of a system formed from 108 particles interacting with 
the CSW potential (filled circles), as obtained from MD simulations,  and with the SW potential (open circles), taken from Michels and Trappeniers,\cite{michels1982molecular} at several temperatures. In all cases, $\lambda=1.5 \,\sigma$.}
\label{fig_D_108}
\end{figure}

At this point, we have shown that it is possible to accurately predict the self-diffusion coefficient of SW molecules using the CSW intermolecular potential. However, since the self-diffusivity is system-size dependent, it is also useful to estimate $D$ using different number of molecules. This will allow to extrapolate $D$ to the limit of thermodynamic limit ($N \to \infty$).\cite{maginn2019best}
Unfortunately, this requires to perform additional simulations. Instead of this, we choose an alternative approach that involves simulating a system formed by $1000$  molecules and the use of the Yeh and Hummer\cite{yeh2004system} correction given by:

\begin{equation}
    D_{\infty} = D_{sim} + \frac{k_{B}\,T\, \xi}{6 \, \pi\, \eta \,L},
    \label{eq_d_inf}
\end{equation}

\noindent 
where $D_{\infty}$ is the self-diffusivity in the infinite system, $D_{sim}$ is the self-diffusion calculated simulating a system with $N$ particles in a cubic box of side $L$, $k_{B}$ is the Boltzmann constant, $T$ the temperature, $\xi=2.837298$, and $\eta$ the shear viscosity. Note that we use in Eq.~\eqref{eq_d_inf} the estimated values of $\eta$ obtained using the CSW intermolecular potential as it is explained in the following Section (all the self-diffusion coefficient values obtained in this work for systems formed by $108$ and $1000$ molecules, as well as the value obtained using Eq.~\eqref{eq_d_inf}, are included in the Supporting information).

Fig.~\ref{fig_D_1000} shows the self-diffusion coefficient, as a function of density, at supercritical and subcritical temperatures as obtained from MD-NVT simulations of systems formed by $1000$ molecules that interact via the CSW potential. As can be seen, the values obtained simulating $1000$ molecules are slightly higher than those obtained using $108$ molecules, as expected. For this reason we do not compare both sets of data in a single plot. As in Fig.~\ref{fig_D_108}, we also use a logarithmic $D^{*}T^{*}-\rho^{*}$ representation in order to better show the behavior of the self-diffusion coefficient as a function of density. As can be seen, D behaves as a monotonically decreasing function of density and shows a change in the curvature for $\rho^{*} < 0.2$. It is remarkable that the same change in curvature is also predicted in the work of Michels and Trappeniers.\cite{michels1982molecular}

\begin{figure}
\includegraphics[width=\columnwidth]{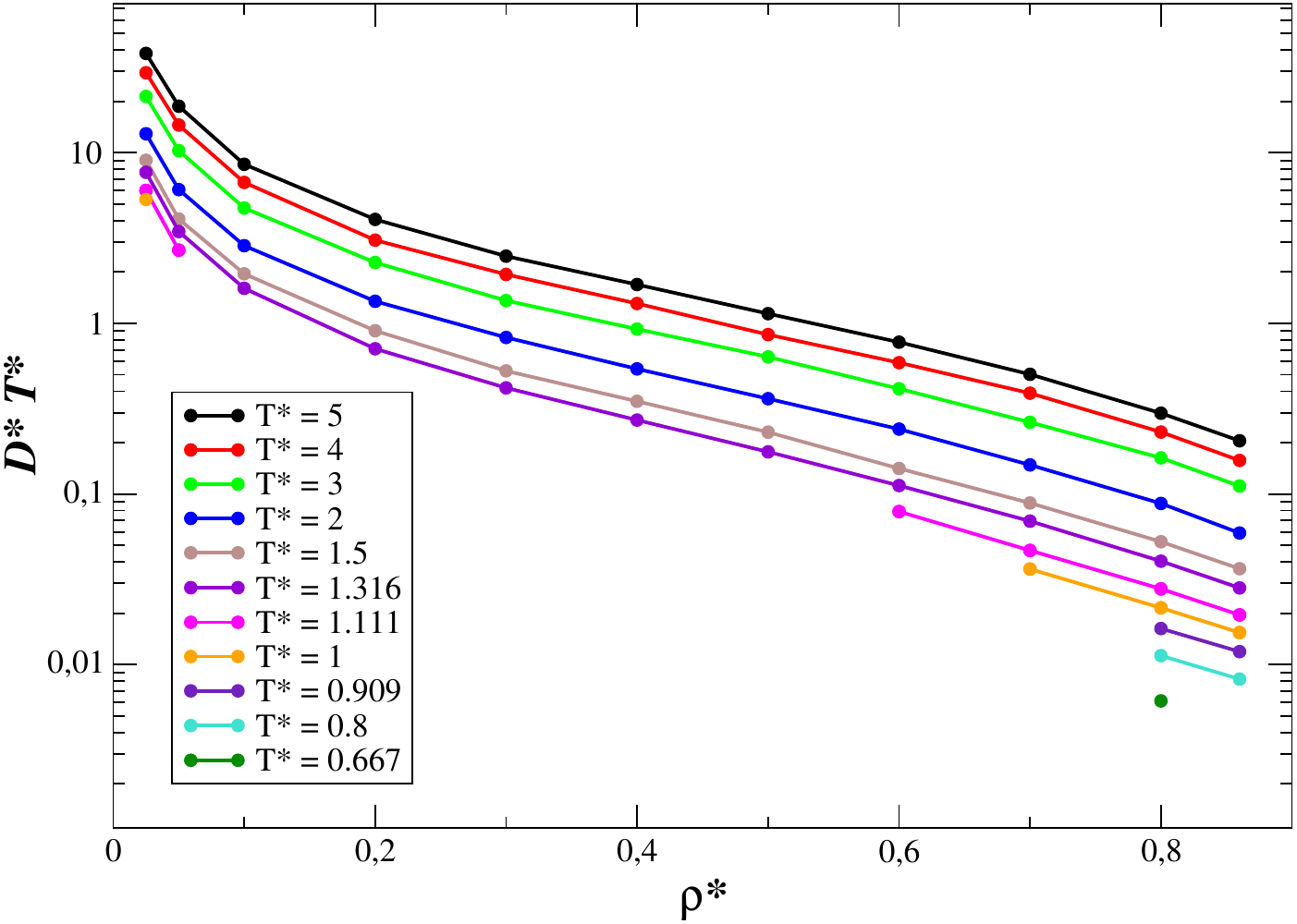}
\caption{Self-diffusion coefficient $D^{*}$, as a function of density, $\rho^{*}$, of a system formed from 1000 particles interacting with 
the CSW potential, as obtained from MD simulations at several temperatures. In all cases, $\lambda=1.5 \,\sigma$.}
\label{fig_D_1000}
\end{figure}

\subsection{Shear-viscosity}

We have also considered the behavior of the shear-viscosity, as a function of density, of spherical SW molecules in the same range of temperatures previously studied. To this end, we use the Green-Kubo formula given by Eq.~\eqref{eq_eta}. This requires the calculation the stress tensor throughout MD-NVT simulations saving information for short periods of time to correctly capture the behavior of the auto-correlation function in Eq.~\eqref{eq_eta}. According to the literature,\cite{maginn2019best} it is recommendable to save the components of the pressure tensor each $5$ to $10\,\text{fs}$ for systems interacting through continuous intermolecular potentials. However, we have checked that systems of spherical SW molecules that interaction via the CSW intermolecular potential need to save the components of the pressure tensor every $0.5\,\text{fs}$.

It is possible to determine the diagonal, $P_{xx}$, $P_{yy}$, $P_{zz}$, and off-diagonal,  $P_{xy}$, $P_{xz}$, $P_{yz}$, components of the pressure tensor from the same simulations performed to determine the self-diffusion coefficients. In particular, we average the auto-correlation function of several components of the pressure tensor, the three off-diagonal components and the equivalent terms $(P_{xx}- P_{yy})/2$ and $(P_{yy}- P_{zz})/2$.\cite{gonzalez2010shear} 

We use a home-made program to analyze and calculate the shear viscosity curves, as functions of density, at several subcritical and supercritical temperatures. As for the case of diffusivity, the reported data (averages and uncertainties) correspond to the average of $5$ independent runs for each thermodynamic state studied. In each simulation, the value of the viscosity corresponds to the time-average value of $\eta(t)$ from times at which the plateau is reached until the upper time limit in Eq.~\eqref{eq_eta}. In all cases, this time limit is fixed at $100\,\text{ps}$. The time at which the plateau occurs depends on the thermodynamic conditions: $\sim35\,\text{ps}$, for low temperatures and densities, $\sim20\,\text{ps}$ at low temperatures and high densities or vice versa, and $\sim8\,\text{ps}$ for high temperatures and densities. It is interesting to mention that it is possible to estimate each value of $\eta$ in a different manner. Particularly, one could fit the data using some analytic function, as suggested by Maginn \emph{et al.},~\cite{maginn2019best} and to obtain results close to those obtained using in this work.

Fig.~\ref{fig_eta} shows the shear viscosity, as a function of density, at different temperatures as obtained from MD-NVT simulation of systems formed by $108$ and $1000$ molecules interacting through the CSW intermolecular potential with range $\lambda=1.5\,\sigma$. We have also included the simulation data obtained by Michels and Trappeniers.~\cite{michels1980molecular,michels1980-2molecular} As can be seen, agreement between results obtained using the standard SW potential and the CSW implementation of Zer\'on~\cite{zeron2018continuous} is excellent in the whole range conditions of temperature and density. Particularly, $\text{AAD}\%\sim 3\%$ for simulations using $108$ molecules. The simulation results obtained using $108$ and $1000$ molecules are very similar, indicating that shear-viscosity is a collective property that it does not depend on the system size. A detailed account of the results obtained in this section are tabulated and included in the Supplementary Material.

\begin{figure}
\includegraphics[width=\columnwidth]{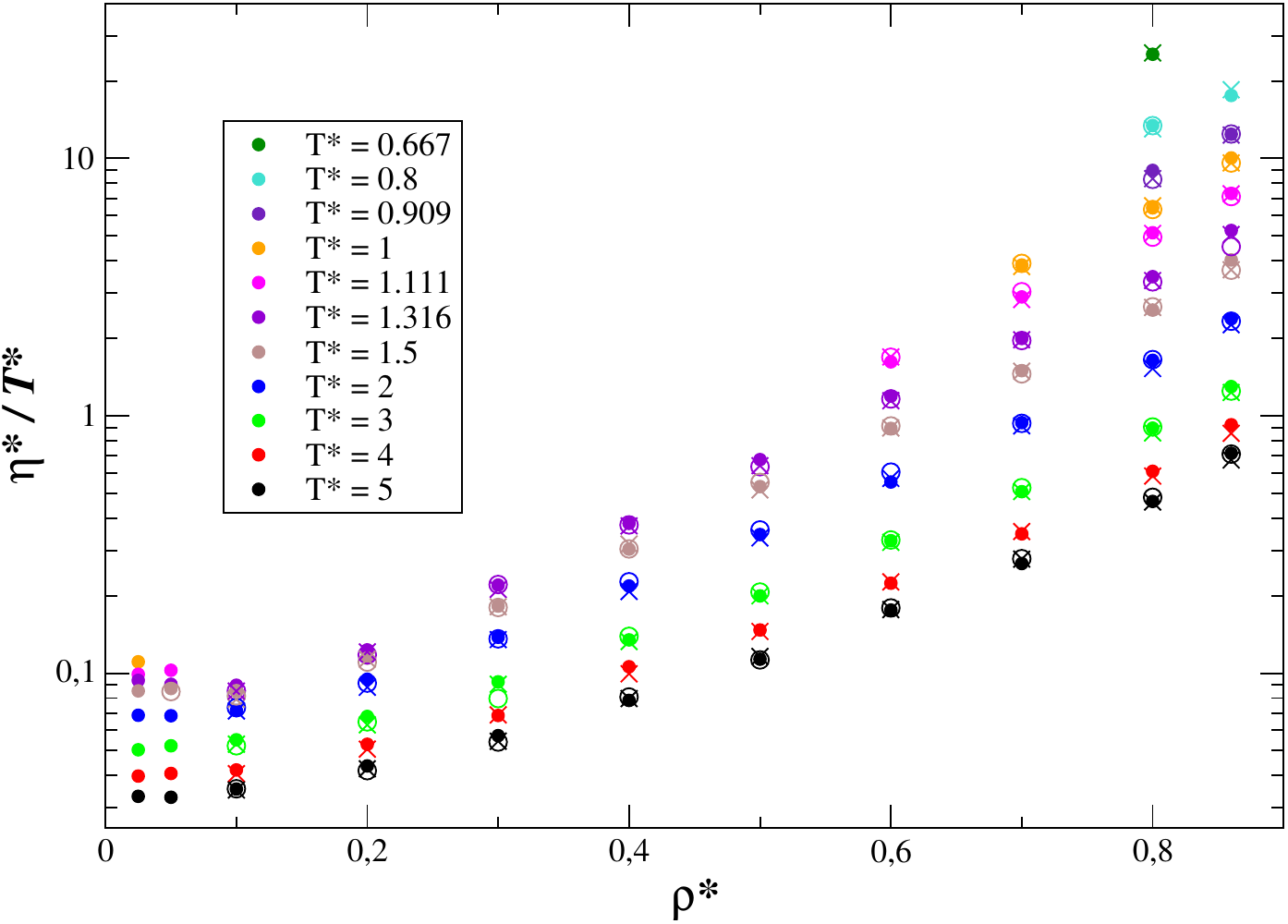}
\caption{Shear-viscosity, $\eta^{*}$, as a function of density, $\rho^{*}$, of systems formed from 108 (crosses) and 1000 particles (filled circles) interacting with the CSW potential at several temperatures. We have also included the results corresponding to a system formed from 108 particles (open circles) interacting with the SW potential at the same temperature.\cite{michels1980molecular,michels1980-2molecular} In all cases, $\lambda=1.5 \,\sigma$.}
\label{fig_eta}
\end{figure}

\subsection{Equilibrium properties}

We now concentrate in equilibrium magnitudes, including structural properties, of the SW fluid determined using the CSW version of the intermolecular potential. Particularly, we focus on internal energy, pressure, and radial distribution function. Note that simulation data in the literature is less scarce in this case since equilibrium properties can be routinely computed using standard MC simulation codes. The reason is that only the intermolecular potential, and not the intermolecular forces (that exhibit discontinuities), are needed to evaluate properties using standard MC moves.

Equilibrium properties of the SW fluid interacting via its CSW version have been previously investigated by Zer\'on \emph{et al.}~\cite{zeron2018continuous} However, the previous study has focused on the behavior of the system at a reduced number of temperatures and densities. Taking advantage of the simulations performed to predict transport properties of system formed by $1000$ molecules, with potential range $\lambda =1.5 \,\sigma$, we have calculated the internal energy, compressibility factor, and radial distribution function at all states analyzed in this work. This allows to compare equilibrium properties of the SW fluid obtained from MC simulations using the standard SW potential and those obtained from MD-NVT simulations using the CSW version of the intermolecular potential.
 
The compressibility factor, $Z$, that measures the non-ideality a fluid system ($Z=1$), is defined as:
\begin{equation}
    Z = \frac{P\,V}{N\,k_{B}\,T} = \frac{P^{*}}{\rho^{*}\,T^{*}}\,
\end{equation}
where $P$ is the pressure, $V$ the volume, $k_{B}$ the Boltzmann constant, and $T$ the temperature of the system. Fig.~\ref{fig_z} shows the compressibility factor, as a function of the density, in a wide range of temperatures for SW fluids interacting via the CSW version of the potential with range $\lambda =1.5 \,\sigma$. We have also included in the plot MC simulation data taken from the literature.~\cite{largo2003generalized,henderson1976monte,labik1999sp,gil1995structure,michels1980molecular} As we can see, there is a good agreement between the estimations obtained using both versions of the SW intermolecular potential. Particularly, we find that $\text{AAD}\% <1\%$.

\begin{figure}
\includegraphics[width=\columnwidth]{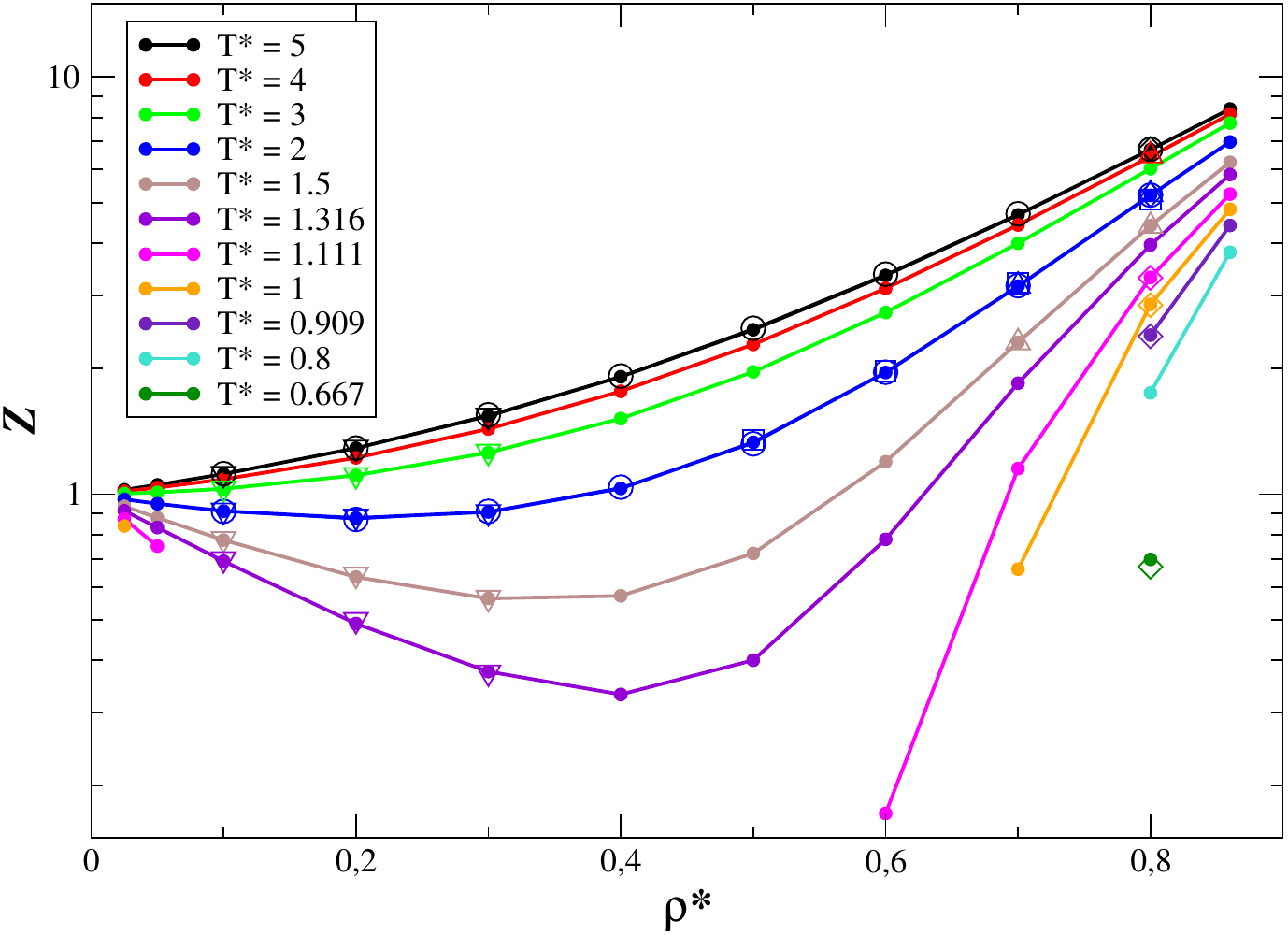}
\caption{Compressibility factor, $Z$, as a function of density, at several temperatures. Filled circles are the MD predictions using the CSW potential and open symbols are the MC data of Largo and Solana\cite{largo2003generalized}(circles), Henderson \emph{et al.}\cite{henderson1976monte} (squares), Labik \emph{et al.}\cite{labik1999sp} (diamond), Gil-Villegas\emph{et al.}\cite{gil1995structure} (triangle up), and DMD predictions of Michels and Trappeniers\cite{michels1980molecular} (triangle down) using the SW potential. In all cases, $\lambda =1.5 \,\sigma$. The continuous curves are guides to the eye.}
\label{fig_z}
\end{figure}

We have also obtained the internal energy, as a function of density, in a wide range of temperatures. Isotherms for the reduced excess internal energy are represented in Fig.~\ref{fig_u} obtained using both versions of the SW potential with a range $\lambda=1.5 \,\sigma$. Again, agreement between results obtained using both intermolecular potentials is excellent in the whole range of temperatures and densities ($\text{AAD}\%\sim 0.3\%$). The numerical data of $u^{*}$ and $Z$ for the system formed from $1000$ molecules interacting via the CSW version of the potential are tabulated in the Supporting Information.

\begin{figure}
\includegraphics[width=\columnwidth]{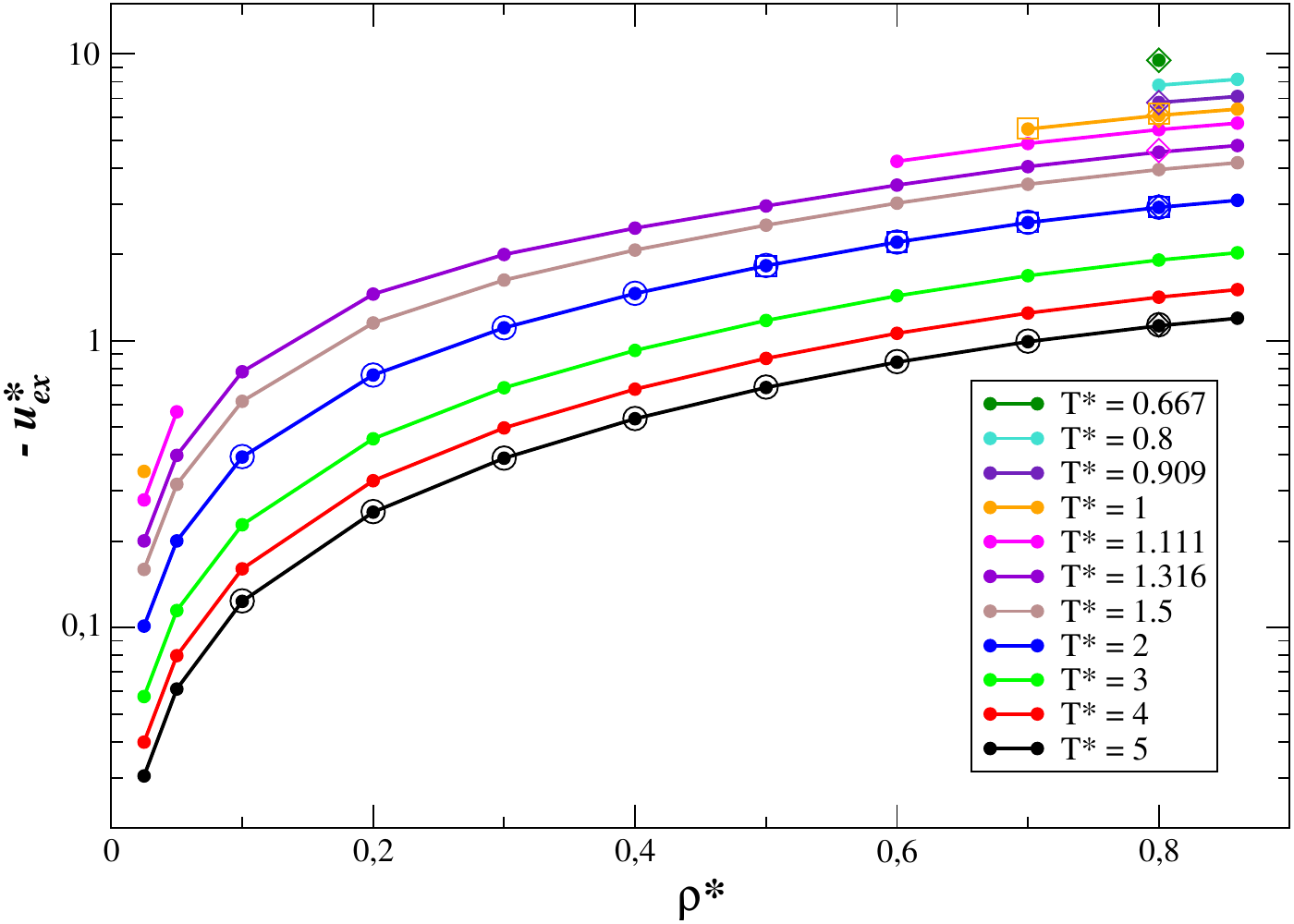}
\caption{Reduced internal energy of excess, $u_{ex}^{*}= U_{ex}/Nk_{B}T$, as a function of density, $\rho^{*}$, at several temperatures. The meaning of the symbols and curves are the same as in Fig.~\ref{fig_z}.}
\label{fig_u}
\end{figure}

Finally, we have determined two structural properties of the SW fluid. Particularly, we analyze the radial distribution function of the fluid. As a representative example, here we only show three $g(r^{*})$ curves at low, intermediate, and high densities and different temperatures. MD-NVT simulation predictions obtained using the CSW version of the intermolecular potential are compared with data taken from the literature~\cite{largo2005pair,henderson1976monte} as obtained using MC simulations using the standard SW potential with range $\lambda = 1.5 \,\sigma$. As can be seen in in Fig.~\ref{fig_rdf}, agreement between MC and MD simulation data is excellent in all cases.

The coordination number of molecules in a fluid can be easily computed from the radial distribution function. This property, $N_{c}$, accounts for the average number of neighbor particles within the well.  $N_{c}$ can be readily calculated through the following expression:

\begin{equation}
    N_{c}= 4\,\pi\,\rho^{*}\int_{1}^{\lambda}\,g(r^{*}) \, r^{*^{2}} dr^{*}
\end{equation}

It is interesting to mention that there exist several theoretical models developed in the literature that allow to estimate the coordination number of SW fluids.\cite{lee1985generalized,heyes1991coordination,largo2002theory,nasrifar2003equation,haghtalab2009new} As can be observed in Fig.~\ref{fig_nc}, agreement between MC simulation data (using the standard SW potential) taken from the literature and the results obtained in this work using the CSW version of the intermolecular potential is excellent in all cases ($\text{AAD}\sim 0.4\%$). See Supplementary Material for the complete set of $N_{c}$ values calculated in this work.

\begin{figure}
\includegraphics[width=\columnwidth]{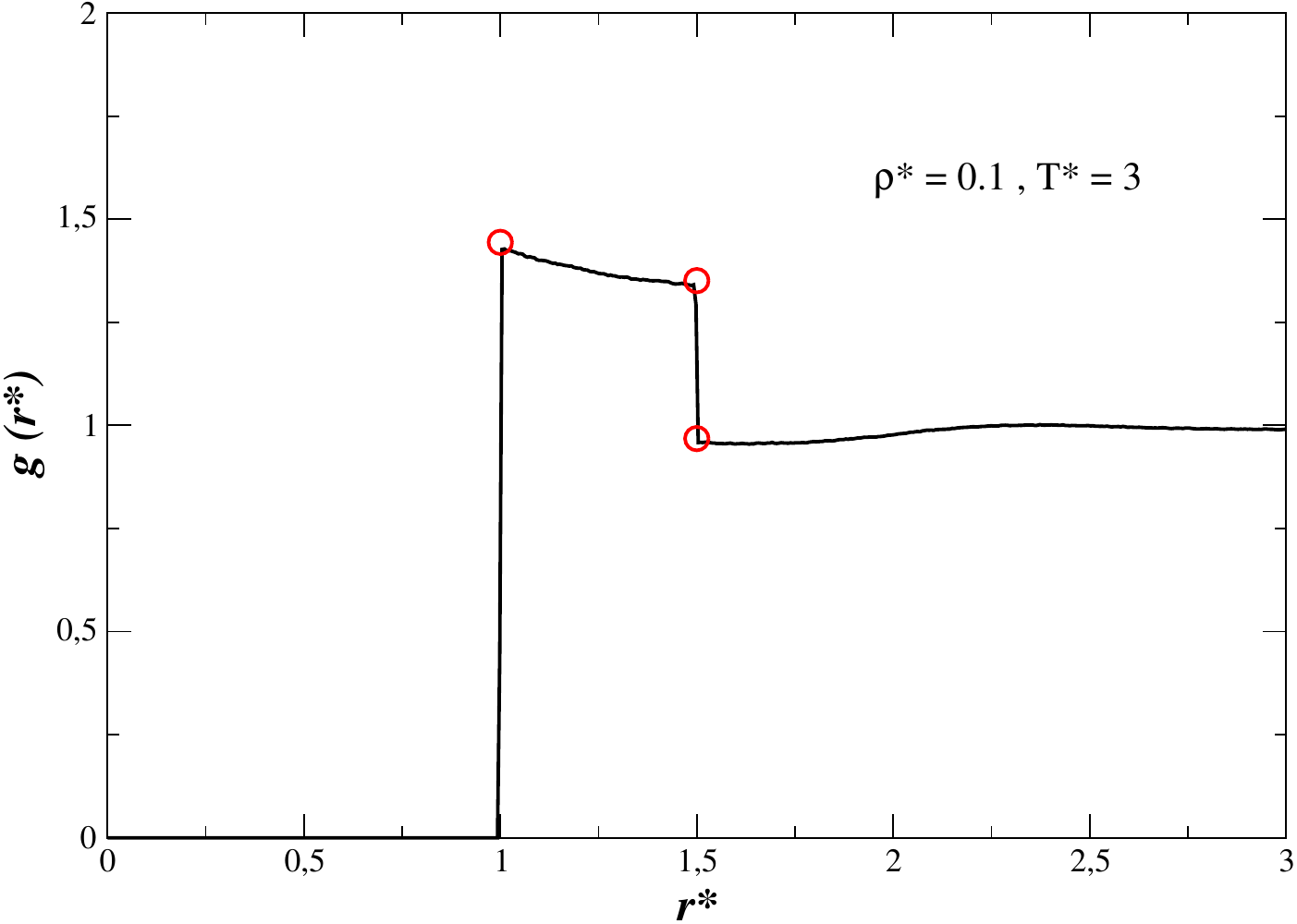}
\includegraphics[width=\columnwidth]{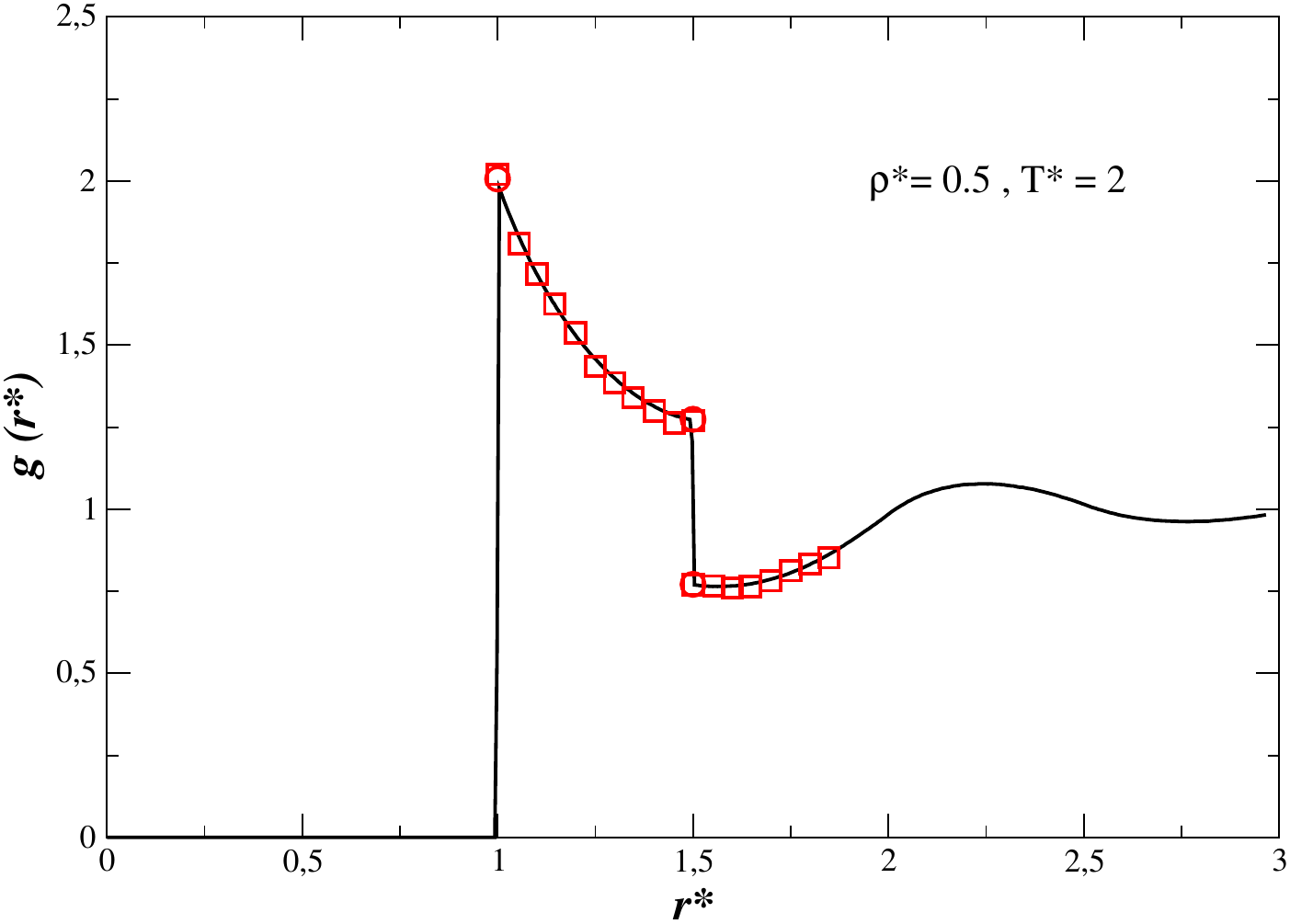}
\includegraphics[width=\columnwidth]{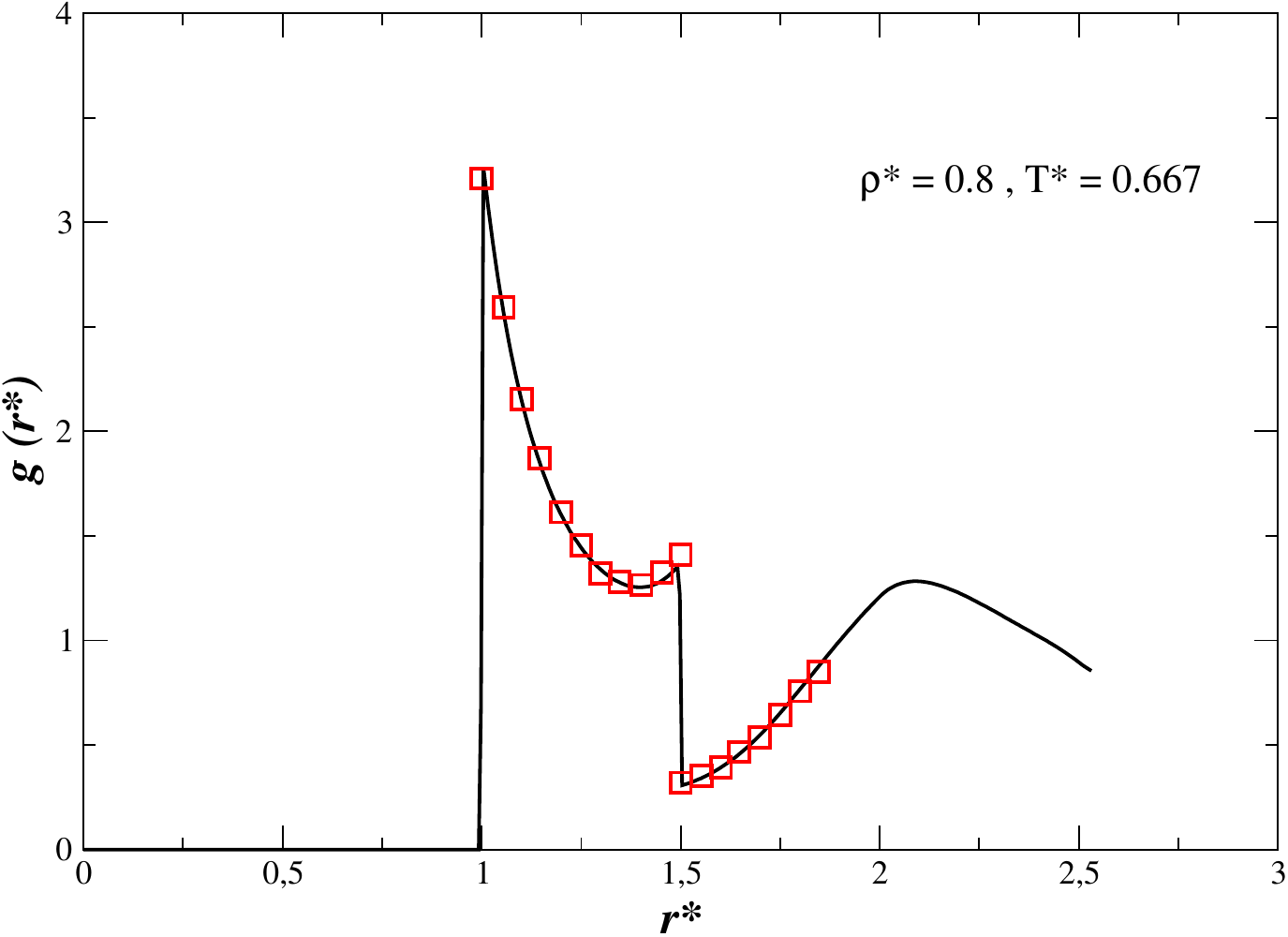}
\caption{Radial distribution function, $g(r^{*})$, as a function of reduced distance, $r^{*} \equiv r/\sigma$, at different densities and temperatures. Open circles represent the MC data using the SW potential taken from the the work of Largo \emph{et al.}\cite{largo2005pair} and the open squares from the work of Henderson \emph{et al.}\cite{henderson1976monte} The continuous line are obtained from MD simulations using the CSW potential. In all cases, $\lambda=1.5 \,\sigma$.}
\label{fig_rdf}
\end{figure}

\begin{figure}
\includegraphics[width=\columnwidth]{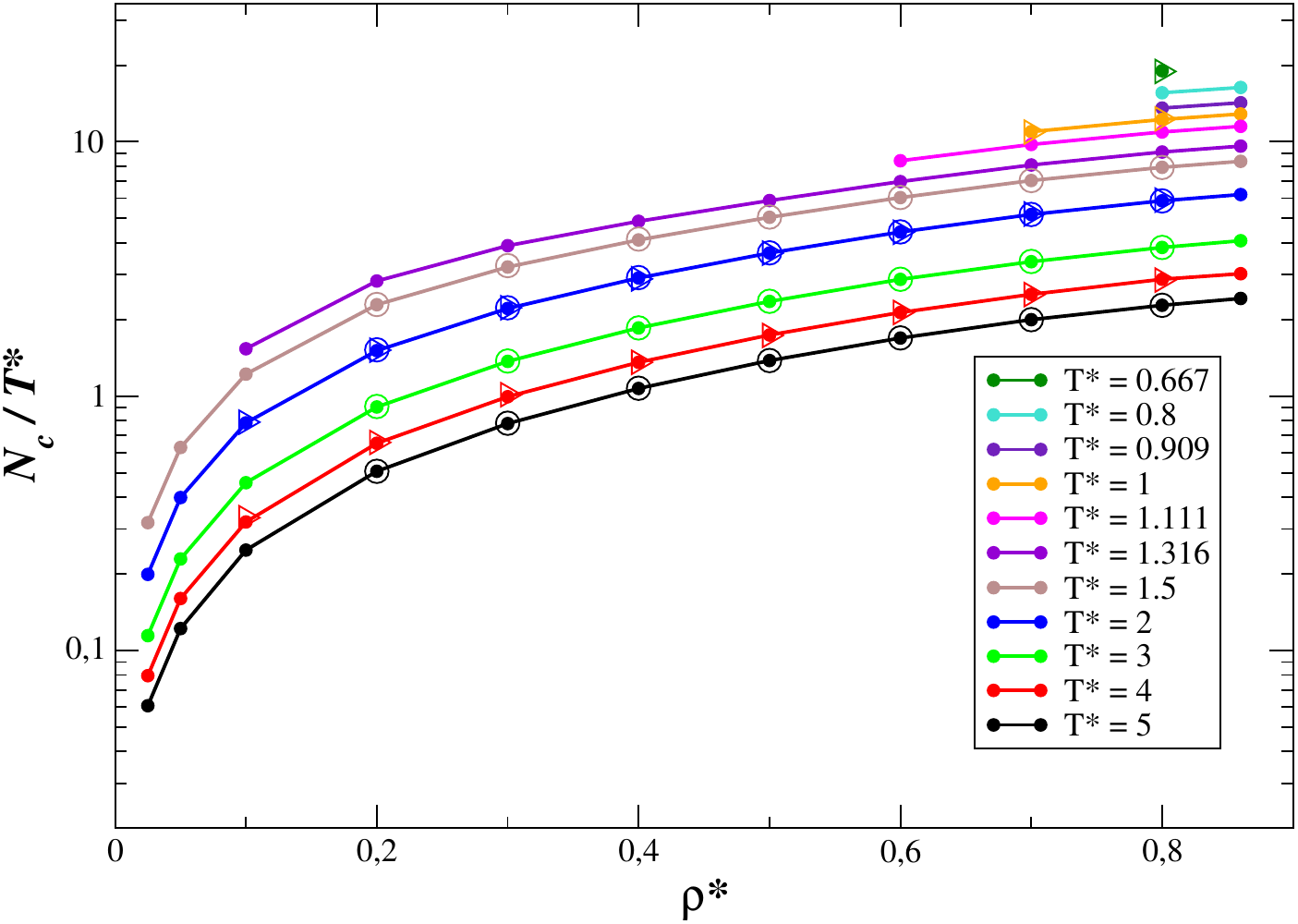}
\caption{Coordination number, as a function of reduced density, $\rho^{*}$, along several isotherms. Filled circles are results obtained from MD simulations using the CSW potential, open circles are MC data using the SW potential from Largo and Solana,\cite{largo2002theory} and triangles right from Lee \emph{et al.}\cite{lee1985generalized} In all cases, $\lambda = 1.5 \,\sigma$.}
\label{fig_nc}
\end{figure}

\section{Conclusions}

We have performed MD simulations in the NVT or canonical ensemble of systems formed from $108$ and $1000$ molecules interacting through the continuous square-well potential with range $\lambda = 1.5 \,\sigma$. Conditions at which simulations are performed correspond to the gas and liquid phases, including subcritical and supercritical regions of the phase diagram. Transport properties (shear viscosity and self-diffusion coefficient), as well as equilibrium (internal energy and compressibility factor) and structural properties (radial distribution function and coordination number) have been determined in a wide range of temperatures and pressures. We have compared the results obtained using the combination of MD simulations and the continuous version of the square-well potential with MC simulation data obtained using the standard SW potential taken from the literature.

The absolute average relative deviation ($\text{AAD\%}$) between our results and those taken from the literature found for transport properties is $3\%$, approximately, for self-diffusion coefficient and shear viscosity. Agreement in the case of equilibrium properties is even better, with AAD\% is below $1\%$. It is important to recall here that we have not observed a trend to systematically subestimate or overestimate data, which corroborated that deviations in predictions are part of the intrinsic uncertainty associated to the simulations.

The main conclusion of this work is that the continuous version of the SW potential proposed by Zer\'on \emph{et al.}~\cite{zeron2018continuous} is able to reproduce accurately not only equilibrium and structural, but also transport properties of the SW fluid with potential range $\lambda=1.5\,\sigma$ in a wide range of densities and temperatures.


\section*{Acknowledgements}
This work was finnanced by Ministerio de Ciencia e Innovaci\'on (Grant No.~PID2021-125081NB-I00), Junta de Andalucía (P20-00363), and Universidad de Huelva (P.O. FEDER UHU-1255522 and FEDER-UHU-202034), all four cofinanced by EU FEDER funds. We also acknowledge the Centro de Supercomputaci\'on de Galicia (CESGA, Santiago de Compostela, Spain) for providing access to computing facilities. We also acknowledge RES resources provided by Barcelona Supercomputing Center in Mare Nostrum to FI-2023-2-0041 and FI-2023-3-0011.

%

\bibliography{bibfjblas}

\end{document}


\title{Supporting Information\\ Transport properties of the square-well fluid from molecular dynamics simulation}

\author{Iv\'an M. Zer\'on}
\affiliation{Laboratorio de Simulaci\'on Molecular y Qu\'imica Computacional, CIQSO-Centro de Investigaci\'on en Qu\'imica Sostenible and Departamento de Ciencias Integradas, Universidad de Huelva, 21006 Huelva Spain}

\author{Marina Cueto-Mora}
\affiliation{Laboratorio de Simulaci\'on Molecular y Qu\'imica Computacional, CIQSO-Centro de Investigaci\'on en Qu\'imica Sostenible and Departamento de Ciencias Integradas, Universidad de Huelva, 21006 Huelva Spain}

\author{Felipe J. Blas}
\affiliation{Laboratorio de Simulaci\'on Molecular y Qu\'imica Computacional, CIQSO-Centro de Investigaci\'on en Qu\'imica Sostenible and Departamento de Ciencias Integradas, Universidad de Huelva, 21006 Huelva Spain}
\email{felipe@uhu.es}

\maketitle

\section{Introduction}

Numerical data of all properties analyzed in the main paper for systems of 108 and 1000 particles which interact with the continuous square well (CSW) potential of range $\lambda = 1.5$ are tabulated in this material.

Values in this work will be presented in reduced units: density $\rho^{*}= \rho \, \sigma^{3}$, temperature $T^{*} = k_{B}T/\epsilon$, excess internal energy per particle $u_{ex}^{*} = U_{ex}/(N\epsilon)$, pressure $P^{*}=P\sigma^{3}/\epsilon$, compressibility factor $Z=PV/(Nk_{B}T)=P^{*}/(\rho^{*}T^{*})$, self-diffusion coefficient $D^{*}= D \sqrt{m/(\sigma^{2}k_{B}T)}$ , shear-viscosity $\eta^{*} =\eta \sigma^{2}/ \sqrt{(mk_{B}T)}$ \,.

%


\begin{table}[htb]
    \rotatebox{90}{
    \begin{minipage}{1.22\textwidth}
    \centering
    \begin{tabular}{||c||c|c|c|c|c|c|c|c|c|c|c|}
\hline
\hline
\backslashbox{$\rho^{*}$}{$T^{*}$}& 5  & 4 & 3 & 2    & 1.5       & 1.316      & 1.111      & 1          & 0.909      & 0.8        & 0.666 \\
\hline
\hline
0.1 & 1.73(19)  & 1.71(29)  & 1.482(98) & 1.398(71) & 1.272(81) & 1.14(14)  & - -       & - -      & - -      & - -      & - -      \\  
0.2 & 0.846(105)& 0.745(59) & 0.670(68) & 0.612(19) & 0.564(21) & 0.509(34) & - -       & - -      & - -      & - -      & - -      \\
0.3 & 0.467(11) & 0.446(50) & 0.426(49) & 0.379(24) & 0.319(29) & 0.305(22) & - -       & - -      & - -      & - -      & - -      \\
0.4 & 0.309(46) & 0.321(23) & 0.268(23) & 0.239(22) & 0.214(29) & 0.193(13) & - -       & - -      & - -      & - -      & - -      \\
0.5 & 0.200(15) & 0.195(30) & 0.191(11) & 0.164(6)  & 0.137(10) & 0.127(8)  & - -       & - -      & - -      & - -      & - -      \\
0.6 & 0.143(5)  & 0.127(6)  & 0.121(6)  & 0.103(13) & 0.082(8)  & 0.080(7)  & 0.060(03) & - -      & - -      & - -      & - -      \\
0.7 & 0.086(9)  & 0.084(6)  & 0.081(7)  & 0.062(6)  & 0.050(3)  & 0.044(1)  & 0.039(04) & 0.033(2) & - -      & - -      & - -      \\
0.8 & 0.055(3)  & 0.053(7)  & 0.046(3)  & 0.039(2)  & 0.029(3)  & 0.028(2)  & 0.022(02) & 0.018(2) & 0.016(1) & 0.012(1) & 0.008(1) \\
0.86& 0.038(4)  & 0.033(3)  & 0.033(3)  & 0.027(3)  & 0.020(2)  & 0.019(3)  & 0.014(01) & 0.013(1) & 0.010(2) & 0.008(1) & - -      \\ 
\hline
    \end{tabular}
    \caption{Self-diffusion coefficient $D^{*}$ for square well ($\lambda = 1.5$) fluids as function of reduced density $\rho^{*}$ and temperature $T^{*}$ from MD simulations using  108 particles interacting with the CSW potential of the same range. Number in parenthesis correspond to the uncertainty in the last digit.}
   \end{minipage}}
\end{table}

\begin{table}[htb]
    \rotatebox{90}{
    \begin{minipage}{1.22\textwidth}
    \centering
    \begin{tabular}{||c||c|c|c|c|c|c|c|c|c|c|c|}
\hline
\hline
\backslashbox{$\rho^{*}$}{$T^{*}$}& 5  & 4 & 3 & 2    & 1.5       & 1.316      & 1.111      & 1          & 0.909      & 0.8        & 0.666 \\
\hline
\hline
0.025& 7.64(22)  & 7.36(13)   & 7.13(15)  & 6.46(12)  & 6.03(17)  & 5.84(14)  & 5.42(15) & 5.32(15) & - -      & - -      &          \\
0.05 & 3.747(44) & 3.632(93)  & 3.43(13)  & 3.040(53) & 2.724(78) & 2.626(53) & 2.42(11) & - -      & - -      & - -      &          \\
0.1  & 1.711(39) & 1.677(36)  & 1.583(56) & 1.427(40) & 1.301(47) & 1.218(18) & - -      & - -      & - -      & - -      &          \\
0.2  & 0.814(12) & 0.768(22)  & 0.757(31) & 0.674(31) & 0.603(21) & 0.539(15) & - -      & - -      & - -      & - -      &          \\
0.3  & 0.495(15) & 0.484(17)  & 0.453(12) & 0.414(16) & 0.351(11) & 0.318(7)  & - -      & - -      & - -      & - -      &          \\
0.4  & 0.338(7)  & 0.327(12)  & 0.309(7)  & 0.271(8)  & 0.233(9)  & 0.206(9)  & - -      & - -      & - -      & - -      &          \\
0.5  & 0.228(9)  & 0.215(1)   & 0.212(7)  & 0.181(6)  & 0.154(6)  & 0.134(3)  & - -      & - -      & - -      & - -      &          \\
0.6  & 0.155(3)  & 0.147(6)   & 0.138(4)  & 0.120(4)  & 0.094(3)  & 0.085(5)  & 0.071(3) & - -      & - -      & - -      &          \\
0.7  & 0.101(2)  & 0.098(2)   & 0.088(2)  & 0.074(2)  & 0.059(2)  & 0.053(1)  & 0.042(1) & 0.036(1) & - -      & - -      &          \\
0.8  & 0.060(2)  & 0.058(1)   & 0.054(1)  & 0.044(2)  & 0.035(1)  & 0.031(1)  & 0.025(0) & 0.022(1) & 0.018(1) & 0.014(0) & 0.009(0) \\
0.86 & 0.041(1)  & 0.039(1)   & 0.037(1)  & 0.030(0)  & 0.024(0)  & 0.021(1)  & 0.018(1) & 0.015(1) & 0.013(1) & 0.010(0) & - -      \\
\hline
    \end{tabular}
    \caption{Self-diffusion coefficient $D^{*}$ for square well ($\lambda = 1.5$) fluids as function of reduced density $\rho^{*}$ and temperature $T^{*}$ from MD simulations using  1000 particles interacting with the CSW potential of the same range. Number in parenthesis correspond to the uncertainty in the last digit.}
   \end{minipage}}
\end{table}

\begin{table}[htb]
    \rotatebox{90}{
    \begin{minipage}{1.22\textwidth}
    \centering
    \begin{tabular}{||c||c|c|c|c|c|c|c|c|c|c|c|}
\hline
\hline
\backslashbox{$\rho^{*}$}{$T^{*}$}& 5  & 4 & 3 & 2    & 1.5       & 1.316      & 1.111      & 1          & 0.909      & 0.8        & 0.666 \\
\hline
\hline
0.025& 7.687 & 7.400 & 7.172 & 6.512 & 6.080 & 5.899 & 5.533 & 5.439 &  - -  & - -   & - -   \\
0.05 & 3.800 & 3.686 & 3.486 & 3.105 & 2.791 & 2.700 & 2.534 & - -   &  - -  & - -   & - -   \\
0.1  & 1.751 & 1.718 & 1.625 & 1.475 & 1.356 & 1.278 &  - -  & - -   &  - -  & - -   & - -   \\
0.2  & 0.854 & 0.810 & 0.800 & 0.720 & 0.655 & 0.593 &  - -  & - -   &  - -  & - -   & - -   \\
0.3  & 0.530 & 0.521 & 0.489 & 0.450 & 0.388 & 0.353 &  - -  & - -   &  - -  & - -   & - -   \\
0.4  & 0.366 & 0.353 & 0.336 & 0.296 & 0.257 & 0.228 &  - -  & - -   &  - -  & - -   & - -   \\
0.5  & 0.249 & 0.235 & 0.232 & 0.198 & 0.169 & 0.148 &  - -  & - -   &  - -  & - -   & - -   \\           
0.6  & 0.170 & 0.162 & 0.151 & 0.131 & 0.104 & 0.093 & 0.078 & - -   &  - -  & - -   & - -   \\
0.7  & 0.111 & 0.107 & 0.096 & 0.081 & 0.065 & 0.058 & 0.046 & 0.040 &  - -  & - -   & - -   \\               
0.8  & 0.066 & 0.063 & 0.060 & 0.049 & 0.039 & 0.034 & 0.027 & 0.024 & 0.020 & 0.015 & 0.010 \\
0.86 & 0.045 & 0.043 & 0.041 & 0.033 & 0.027 & 0.024 & 0.019 & 0.017 & 0.014 & 0.011 &       \\
\hline
    \end{tabular}
    \caption{Self-diffusion coefficient at infinite system $D_{\infty}^{*}$ for square well ($\lambda = 1.5$) fluids as function of reduced density $\rho^{*}$ and temperature $T^{*}$ from MD simulations using  1000 particles interacting with the CSW potential of the same range.}
   \end{minipage}}
\end{table}

\begin{table}[htb]
    \rotatebox{90}{
    \begin{minipage}{1.22\textwidth}
    \centering
    \begin{tabular}{||c||c|c|c|c|c|c|c|c|c|c|c|}
\hline
\hline
\backslashbox{$\rho^{*}$}{$T^{*}$}& 5  & 4 & 3 & 2    & 1.5       & 1.316      & 1.111      & 1          & 0.909      & 0.8        & 0.666 \\
\hline
\hline
0.1 & 0.176(9)  & 0.163(5)  & 0.159(9)  & 0.143(5)  & 0.119(4)  & 0.112(7)  & - -       & - -      & - -        & - -       & - -       \\
0.2 & 0.213(6)  & 0.203(9)  & 0.189(6)  & 0.176(5)  & 0.170(10) & 0.159(5)  & - -       & - -      & - -        & - -       & - -       \\
0.3 & 0.272(17) & 0.275(18) & 0.272(7)  & 0.270(11) & 0.272(22) & 0.278(20) & - -       & - -      & - -        & - -       & - -       \\
0.4 & 0.398(22) & 0.398(16) & 0.398(20) & 0.415(19) & 0.476(24) & 0.492(18) & - -       & - -      & - -        & - -       & - -       \\
0.5 & 0.581(11) & 0.581(44) & 0.599(34) & 0.671(26) & 0.773(28) & 0.842(61) & - -       & - -      & - -        & - -       & - -       \\
0.6 & 0.883(35) & 0.905(21) & 0.967(52) & 1.142(33) & 1.346(56) & 1.506(59) & 1.882(25) & - -      & - -        & - -       & - -       \\
0.7 & 1.387(84) & 1.420(91) & 1.524(38) & 1.828(43) & 2.236(82) & 2.60(11)  & 3.14(15)  & 3.80(16) & - -        & - -       & - -       \\
0.8 & 2.314(75) & 2.34(12)  & 2.57(17)  & 3.052(81) & 3.95(22)  & 4.42(32)  & 5.68(35)  & 6.54(10) & 7.59 (37)  & 10.43(43) & 17.14(65) \\
0.86& 3.367(53) & 3.42(10)  & 3.70(18)  & 4.52(24)  & 5.54(33)  & 6.66(9)   & 8.10(41)  & 9.60(63) & 11.20(16)  & 14.79(47) & - -       \\
\hline
    \end{tabular}
    \caption{Shear viscosity $\eta^{*}$ for square well ($\lambda = 1.5$) fluids as function of reduced density $\rho^{*}$ and temperature $T^{*}$ from MD simulations using  108 particles interacting with the CSW potential of the same range. Number in parenthesis correspond to the uncertainty in the last digit.}
   \end{minipage}}
\end{table}

\begin{table}[htb]
    \rotatebox{90}{
    \begin{minipage}{1.22\textwidth}
    \centering
    \begin{tabular}{||c||c|c|c|c|c|c|c|c|c|c|c|}
\hline
\hline
\backslashbox{$\rho^{*}$}{$T^{*}$}& 5  & 4 & 3 & 2    & 1.5       & 1.316      & 1.111      & 1          & 0.909      & 0.8        & 0.666 \\
\hline
\hline
0.025& 0.166(6)  & 0.159(8)  & 0.151(7)  & 0.137(7)  & 0.128(4)  & 0.123(6)  & 0.110(6)  &  0.111(4)  &  - -     &  - -       &  - -      \\
0.05 & 0.165(4)  & 0.163(5)  & 0.157(8)  & 0.137(10) & 0.131(9)  & 0.119(3)  & 0.114(6)  &  - -       &  - -     &  - -       &  - -      \\
0.1  & 0.177(12) & 0.168(8)  & 0.165(6)  & 0.143(6)  & 0.128(5)  & 0.118(5)  &  - -      &  - -       &  - -     &  - -       &  - -      \\
0.2  & 0.218(4)  & 0.212(10) & 0.204(7)  & 0.189(5)  & 0.172(4)  & 0.162(5)  &  - -      &  - -       &  - -     &  - -       &  - -      \\
0.3  & 0.286(20) & 0.274(27) & 0.278(13) & 0.276(6)  & 0.273(11) & 0.29(3)   &  - -      &  - -       &  - -     &  - -       &  - -      \\
0.4  & 0.392(29) & 0.424(15) & 0.404(12) & 0.436(17) & 0.455(27) & 0.503(23) &  - -      &  - -       &  - -     &  - -       &  - -      \\
0.5  & 0.566(14) & 0.589(22) & 0.599(17) & 0.691(31) & 0.796(19) & 0.889(35) &  - -      &  - -       &  - -     &  - -       &  - -      \\
0.6  & 0.879(46) & 0.895(16) & 0.982(34) & 1.106(73) & 1.34(5)   & 1.569(63) & 1.801(76) &  - -       &  - -     &  - -       &  - -      \\
0.7  & 1.333(31) & 1.391(67) & 1.521(74) & 1.88(11)  & 2.25(8)   & 2.63(16)  & 3.22(18)  &  3.86(14)  &  - -     &  - -       &  - -      \\
0.8  & 2.32(9)   & 2.43(9)   & 2.69(10)  & 3.29(17)  & 3.86(16)  & 4.57(26)  & 5.72(29)  &  6.43(20)  & 8.16(62) &  10.73(34) & 16.93(73) \\
0.86 & 3.58(21)  & 3.69(4)   & 3.90(17)  & 4.77(21)  & 6.05(28)  & 6.91(37)  & 8.12(19)  &  10.05(59) & 11.28(7) &  14.04(69) &  - -      \\
\hline
    \end{tabular}
    \caption{Shear viscosity $\eta^{*}$ for square well ($\lambda = 1.5$) fluids as function of reduced density $\rho^{*}$ and temperature $T^{*}$ from MD simulations using  1000 particles interacting with the CSW potential of the same range. Number in parenthesis correspond to the uncertainty in the last digit.}
   \end{minipage}}
\end{table}

\begin{table}[htb]
    \rotatebox{90}{
    \begin{minipage}{1.22\textwidth}
    \centering
    \begin{tabular}{||c||c|c|c|c|c|c|c|c|c|c|c|}
\hline
\hline
\backslashbox{$\rho^{*}$}{$T^{*}$}& 5  & 4 & 3 & 2    & 1.5       & 1.316      & 1.111      & 1          & 0.909      & 0.8        & 0.666 \\
\hline
\hline
0.025& 1.0249(1) & 1.0170(1) & 1.0032(1)  & 0.9724(2) & 0.9358(4) & 0.9127(2)  & 0.8722(3)  & 0.8389(6)  &  - -       &  - -       &  - -     \\
0.05 & 1.0532(1) & 1.0374(1) & 1.0099(1)  & 0.9487(2) & 0.8778(4) & 0.8320(3)  & 0.7515(4)  &  - -       &  - -       &  - -       &  - -     \\   
0.1  & 1.1180(1) & 1.0860(2) & 1.0306(1)  & 0.9113(1) & 0.7762(6) & 0.6910(7)  &  - -       &  - -       &  - -       &  - -       &  - -     \\   
0.2  & 1.2892(1) & 1.2225(3) & 1.1096(3)  & 0.8769(2) & 0.6333(3) & 0.4896(3)  &  - -       &  - -       &  - -       &  - -       &  - -     \\   
0.3  & 1.5396(4) & 1.4339(2) & 1.2576(3)  & 0.9069(3) & 0.5625(3) & 0.3758(3)  &  - -       &  - -       &  - -       &  - -       &  - -     \\   
0.4  & 1.9127(5) & 1.7641(3) & 1.5174(2)  & 1.0326(3) & 0.5708(8) & 0.3311(6)  &  - -       &  - -       &  - -       &  - -       &  - -     \\   
0.5  & 2.4781(4) & 2.2850(2) & 1.9642(4)  & 1.3317(4) & 0.7213(8) & 0.4001(11) &  - -       &  - -       &  - -       &  - -       &  - -     \\   
0.6  & 3.3446(3) & 3.1127(2) & 2.7268(3)  & 1.9574(3) & 1.1964(5) & 0.7793(6)  & 0.1718(11) &  - -       &  - -       &  - -       &  - -     \\   
0.7  & 4.6724(4) & 4.4189(7) & 3.9957(4)  & 3.1517(7) & 2.3121(3) & 1.8459(5)  & 1.1524(6)  & 0.6612(15) &  - -       &  - -       &  - -     \\
0.8  & 6.6942(8) & 6.4433(8) & 6.0281(11) & 5.2058(8) & 4.4009(7) & 3.9588(6)  & 3.3077(7)  & 2.8513(10) & 2.4052(17) & 1.7501(18) & 0.6981(8)\\
0.86 & 8.3792(4) & 8.1440(3) & 7.7548(5)  & 6.9889(8) & 6.2438(9) & 5.8357(8)  & 5.2375(12) & 4.8187(7)  & 4.4068(12) & 3.7991(9)  &  - -     \\
\hline
    \end{tabular}
    \caption{Compressibility Factor $Z$ for square well ($\lambda = 1.5$) fluids as function of reduced density $\rho^{*}$ and temperature $T^{*}$ from MD simulations using 1000 particles interacting with the CSW potential of the same range. Number in parenthesis correspond to the uncertainty in the last digit.}
   \end{minipage}}
\end{table}

\begin{table}[htb]
    \rotatebox{90}{
    \begin{minipage}{1.3\textwidth}
    \centering
    \begin{tabular}{||c||c|c|c|c|c|c|c|c|c|c|c|}
\hline
\hline
\backslashbox{$\rho^{*}$}{$T^{*}$}& 5  & 4 & 3 & 2    & 1.5       & 1.316      & 1.111      & 1          & 0.909      & 0.8        & 0.666 \\
\hline
\hline
0.025& -0.1520(1) & -0.1595(2) & -0.1726(3) & -0.2026(4) & -0.2394(3)  & -0.2639(5)  & -0.3098(9) & -0.3504(14) &  - -       &  - -       &  - -      \\
0.05 & -0.3055(2) & -0.3193(2) & -0.3440(1) & -0.4012(3) & -0.4736(12) & -0.5246(8)  & -0.6283(8) &  - -        &  - -       &  - -       &  - -      \\       
0.1  & -0.6173(2) & -0.6415(2) & -0.6850(3) & -0.7869(3) & -0.9228(11) & -1.0264(17) &  - -       &  - -        &  - -       &  - -       &  - -      \\
0.2  & -1.2640(1) & -1.3005(0) & -1.3658(3) & -1.5198(5) & -1.7317(3)  & -1.9138(12) &  - -       &  - -        &  - -       &  - -       &  - -      \\
0.3  & -1.9484(2) & -1.9882(2) & -2.0585(2) & -2.2210(3) & -2.4430(3)  & -2.6298(14) &  - -       &  - -        &  - -       &  - -       &  - -      \\
0.4  & -2.6738(1) & -2.7122(2) & -2.7781(1) & -2.9245(2) & -3.1096(5)  & -3.2529(9)  &  - -       &  - -        &  - -       &  - -       &  - -      \\
0.5  & -3.4348(1) & -3.4713(2) & -3.5312(1) & -3.6552(3) & -3.7936(2)  & -3.8851(5)  &  - -       &  - -        &  - -       &  - -       &  - -      \\
0.6  & -4.2104(1) & -4.2467(1) & -4.3044(1) & -4.4164(1) & -4.5282(2)  & -4.5920(2)  & -4.6920(2) &  - -        &  - -       &  - -       &  - -      \\
0.7  & -4.9624(1) & -4.9999(0) & -5.0579(1) & -5.1666(1) & -5.2709(1)  & -5.3279(1)  & -5.4120(1) & -5.4711(2)  &  - -       &  - -       &  - -      \\
0.8  & -5.6413(2) & -5.6791(2) & -5.7355(2) & -5.8371(1) & -5.9312(1)  & -5.9819(1)  & -6.0550(0) & -6.1063(1)  & -6.1557(2) & -6.2275(3) & -6.3406(1)\\
0.86 & -5.9937(0) & -6.0310(1) & -6.0850(1) & -6.1790(1) & -6.2640(1)  & -6.3093(2)  & -6.3744(2) & -6.4196(1)  & -6.4634(2) & -6.5274(3) &  - -      \\
\hline
    \end{tabular}
    \caption{Reduced internal energy of excess $u_{ex}^{*}= U_{ex}/N\epsilon$ for square well ($\lambda = 1.5$) fluids as function of reduced density $\rho^{*}$ and temperature $T^{*}$ from MD simulations using 1000 particles interacting with the CSW potential of the same range. Number in parenthesis correspond to the uncertainty in the last digit.}
   \end{minipage}}
\end{table}

\begin{table}[htb]
    \rotatebox{90}{
    \begin{minipage}{1.22\textwidth}
    \centering
    \begin{tabular}{||c||c|c|c|c|c|c|c|c|c|c|c|}
\hline
\hline
\backslashbox{$\rho^{*}$}{$T^{*}$}& 5  & 4 & 3 & 2    & 1.5       & 1.316      & 1.111      & 1          & 0.909      & 0.8        & 0.666 \\
\hline
\hline
0.025&0.303(2) &0.318(2) &0.343(2) &0.398(2) &0.477(3) &0.524(3)       &0.606(3)       &0.688(3)       & - -           & - -           & - -   \\
0.05 &0.609(4) &0.640(4) &0.687(4) &0.798(4) &0.942(4) &1.05(1)        &1.24(1)        & - -           & - -           & - -           & - -   \\
0.1  &1.24(1)  &1.28(1)  &1.37(1)  &1.57(1)  &1.83(1)  &2.02(1)        & - -           & - -           & - -           & - -           & - -   \\
0.2  &2.53(2)  &2.61(2)  &2.72(1)  &3.02(1)  &3.43(1)  &3.73(2)        & - -           & - -           & - -           & - -           & - -   \\
0.3  &3.90(2)  &3.98(2)  &4.11(2)  &4.42(2)  &4.83(2)  &5.14(2)        & - -           & - -           & - -           & - -           & - -   \\
0.4  &5.36(3)  &5.43(3)  &5.56(3)  &5.83(3)  &6.17(3)  &6.41(3)        & - -           & - -           & - -           & - -           & - -   \\
0.5  &6.90(3)  &6.96(4)  &7.07(4)  &7.31(4)  &7.57(3)  &7.73(3)        & - -           & - -           & - -           & - -           & - -   \\
0.6  &8.46(5)  &8.53(4)  &8.63(4)  &8.84(5)  &9.05(4)  &9.18(4)        &9.36(4)        & - -           & - -           & - -           & - -   \\
0.7  &9.99(5)  &10.05(5) &10.13(3) &10.36(5) &10.56(5) &10.67(5)       &10.84(6)       &10.96(5)       & - -           & - -           & - -   \\
0.8  &11.38(6) &11.52(5) &11.53(7) &11.72(6) &11.90(6) &12.00(6)       &12.14(6)       &12.24(6)       &12.34(6)       &12.48(5)       &12.70(5)\\
0.86 &12.11(8) &12.11(8) &12.25(8) &12.41(7) &12.57(7) &12.66(7)       &12.78(6)       &12.87(6)       &12.95(6)       &13.08(7)       & \\
\hline
    \end{tabular}
    \caption{Coordination number $N_{c}$ for square well ($\lambda = 1.5$) fluids as function of reduced density $\rho^{*}$ and temperature $T^{*}$ from MD simulations using  1000 particles interacting with the CSW potential of the same range. Number in parenthesis correspond to the uncertainty in the last digit.}
   \end{minipage}}
\end{table}

\bibliography{bibfjblas}